\documentclass[preprint,english,amsart,preprintnumbers,amsmath,amssymb,floatfix]{revtex4}
\usepackage{tikz,xcolor}
\usepackage[colorlinks = true,  
linkcolor = blue,
urlcolor  = blue,
citecolor = blue,
anchorcolor = blue]{hyperref} 

\definecolor{lime}{HTML}{A6CE39}
\DeclareRobustCommand{\orcidicon}{%
	\begin{tikzpicture}
		\draw[lime, fill=lime] (0,0) 
		circle [radius=0.16] 
		node[white] {{\fontfamily{qag}\selectfont \tiny ID}};
		\draw[white, fill=white] (-0.0625,0.095) 
		circle [radius=0.007];
	\end{tikzpicture}
	\hspace{-2mm}
}

\foreach \x in {A, ..., Z}{%
	\expandafter\xdef\csname orcid\x\endcsname{\noexpand\href{https://orcid.org/\csname orcidauthor\x\endcsname}{\noexpand\orcidicon}}
}

\usepackage[T1]{fontenc}
\usepackage[latin9]{inputenc}
\usepackage{color}
\usepackage{array}
\usepackage{amstext}
\usepackage{graphicx}
\usepackage{esint}

\makeatletter

\@ifundefined{textcolor}{}
{%
 \definecolor{BLACK}{gray}{0}
 \definecolor{WHITE}{gray}{1}
 \definecolor{RED}{rgb}{1,0,0}
 \definecolor{GREEN}{rgb}{0,1,0}
 \definecolor{BLUE}{rgb}{0,0,1}
 \definecolor{CYAN}{cmyk}{1,0,0,0}
 \definecolor{MAGENTA}{cmyk}{0,1,0,0}
 \definecolor{YELLOW}{cmyk}{0,0,1,0}
 }

\@ifundefined{definecolor}
 {\usepackage{color}}{}
\@ifundefined{definecolor}
 {\usepackage{color}}{}
\makeatother

\makeatother

\usepackage{babel}

\begin{document}
%
\newcommand{\MSbar}{\ensuremath{\overline{\text{MS}}\ }}

\title{Charm-quark pole mass from HERA Combined and LHCb charm production data}

\author {A.~Vafaee\orcidA{}}
\affiliation{Department of Physics, Ferdowsi University of Mashhad, P.O.Box 1436, Mashhad, Iran}
\affiliation{School of Particles and Accelerators, Institute for Research in Fundamental Sciences (IPM), P.O.Box 19395-5531, Tehran, Iran}
\author {K.~Javidan\orcidB{}}
\affiliation{Department of Physics, Ferdowsi University of Mashhad, P.O.Box 1436, Mashhad, Iran}
\author {S.~Atashbar Tehrani\orcidC{}}
\affiliation{School of Particles and Accelerators, Institute for Research in Fundamental Sciences (IPM), P.O.Box 19395-5531, Tehran, Iran}


\begin{abstract}
One of the most popular definition for the charm-quark mass is the charm-quark pole mass $m_c^{\rm pole}$. In this contribution, we extract the charm-quark pole mass through perturbative Quantum Chromo Dynamics (pQCD) analysis up to the next-to-next-to-leading order (NNLO) corrections from HERA Combined and LHCb charm production recent data sets. Then, we investigate for the first time the charm-quark pole mass $m_c^{\rm pole}$ pure impact, as an extra pQCD parameter on the proton Parton Distribution Functions (PDFs) at the NNLO corrections.  
\end{abstract}


\maketitle

\section{\label{introduction}Introduction}
The HERA machine as a powerful electron-proton collider study simultaneously neutral current (NC) and charged current (CC) $e^{\pm}p$ collisions and their electroweak unification process. On the other hand, the LHCb detector studies the charm-quark production at the Large Hadron Collider (LHC) in $pp$ reactions at $\sqrt{s} = 7$~TeV.

 At the pQCD level the internal structure of the proton is probed by the experiments known as deep inelastic scattering (DIS) measurements. The DIS experiments serve the central data to determine the nucleon structure in terms of parton distribution functions. Contributions from all active quarks and anti quarks are included by the inclusive neutral NC and CC deep inelastic $e^{\pm}p$ scattering cross sections.
 
 At the DIS measurement level the ratio of the virtual photon couplings corresponding to a heavy quark $Q_h, h=b, c$ are approximated by $f(h) \sim \frac{Q_h^2}{\Sigma{Q_q^2}}$~, where $Q_h=\frac{1}{3},~\frac{2}{3}$ are the $b$-quark and $c$-quark electric charges, respectively and $Q_q$~ with $q=u,d,s,c,b$~ represent the kinematically accessible quark flavors. Accordingly, $f(c) \sim \frac{Q_c^2}{Q_d^2+Q_u^2+Q_s^2+Q_c^2+Q_b^2} = \frac{4}{11} \simeq 0.36$ for the $c$-quark and this means that more than one third (or approximately $36$ percent) of the cross sections come from charm quarks in the final state. This significant contribution of $c$-quark at the HERA events is our main motivation to determine the charm-quark pole mass based on the very recently charm production cross section H1-ZEUS combined (HCC)~\cite{H1:2018flt}, LHCb~\cite{Aaij:2013mga} and HCC$+$LHCb charm production cross section data sets. Then, we investigate the pure impact of the charm-quark pole mass $m_c^{\rm pole}$ as an extra free parameter of the pQCD Lagrangian on the uncertainty bands of gluon distribution and some of its ratios at the NNLO corrections.  

 In this NNLO pQCD analysis we make several fits to exactly separate the role and influence of charm-quark pole mass $m_c^{\rm pole}$ from other phenomenological parameters on the uncertainty bands of PDFs and fit-quality, based on the HCC, LHCb and HCC$+$LHCb data sets within the pQCD framework.
 
 From the pQCD point of view, DIS measurements depend on the various phenomenological input data and knowledge of the PDFs~\cite{Alekhin:2012vu,Gao:2013wwa,Tung:2006tb,Aaron:2009aa,Blumlein:2012bf}. For this reason in addition of charm production cross section from the HERA combined data, the full five LHCb charm production cross section data sets at $\sqrt{s} = 7$~TeV are included to show the sensitivity of the gluon distribution and some of related ratios at low values of $x$, where $x$ is the fraction of proton momentum carried by a parton. Since this kinematic range does not currently covered by other data set, inclusion of the LHCb charm production data at $\sqrt{s} = 7$~TeV dramatically improve the gluon distribution uncertainties and fit quality~\cite{Zenaiev:2016kfl,Zenaiev:2015rfa,Aaij:2013noa,Aaij:2015bpa,Aaij:2016jht,Alves:2008zz,Adinolfi:2012qfa}.

 The outline of this  paper is as follows. In Sec.~(\ref{dis}) we describe the theoretical framework of our study and discuss about the inclusive differential cross section of charm-quark production. We introduce the charm-quark mass in the pQCD approach in Sec.~(\ref{charm mass}). In Sec.~(\ref{methodology}), we describe the data set and our methodology. The results are presented in Sec.~(\ref{results}) and then, we conclude with a summary in Sec.~(\ref{Summary}).

\section{\label{dis}Charm-quark production}
The NC and CC deep inelastic ${e^\pm}p$ scattering at the centre-of-mass energies up to $\sqrt{s} \simeq 320\,$GeV are expressed in terms of the proton generalized structure functions:
\begin{eqnarray}
   \sigma_{r,NC}^{{\pm}}&=&   \frac{d^2\sigma_{NC}^{e^{\pm} p}}{d{x}dQ^2} \frac{Q^4 x}{2\pi \alpha^2 Y_+} = \tilde{F_2} \mp \frac{Y_-}{Y_+} x\tilde{F_3} -\frac{y^2}{Y_+} \tilde{F_{\rm L}}~,                                               
    \label{eq:NC}
\end{eqnarray}
\begin{eqnarray}
\sigma_{r,CC}^{\pm} &=&\frac{2\pi x}{G^2_F} \left[\frac{M^2_W+Q^2}{M^2_W}\right]^2 \frac{d^2\sigma_{CC}^{e^{\pm} p}}{d{x}dQ^2} = \frac{Y_{+}}{2}  W_2^{\pm} \mp \frac{Y_{-}}{2}x  W_3^{\pm} - \frac{y^2}{2} W_L^{\pm}~~,
\label{eq:CC}
\end{eqnarray} 
where $x$ is the Bjorken variable, $y$ is the inelasticity, $Q^2$ is the negative of
four-momentum-transfer squared,  $Y_{\pm} = 1 \pm (1-y)^2$, $\alpha$ is the fine-structure constant which is defined 
at zero momentum transfer and $G_F$ is the Fermi constant~\cite{Vafaee:2017nze}.

 Similarly, the inclusive differential cross section of charm production in DIS is expressed in terms of the dimensionless reduced cross sections:
\begin{eqnarray}
	\sigma_{red}^{C\bar{C}} &=& 
	          \frac{d\sigma^{C\bar{C}}(e^{\pm} p)}{d{x}dQ^2} \frac{Q^4 x}{2 \pi \alpha^2 Y_{+}} = F_2^{C\bar C} \mp \frac{Y_{-}}{Y_{+}}x  F_3^{C\bar C} - \frac{y^2}{Y_{+}}  F_L^{C\bar C}~.
    \label{eq:NCheavy}
\end{eqnarray}
At the low-value regions of $Q^2$ where $Q^2\ll M_Z^2$, the parity-violating structure function, $xF_3$ is neglected and the reduced differential cross section of charm production can be expressed by: 
\begin{eqnarray}
	\sigma_{red}^{C\bar{C}} &=& 
	          \frac{d\sigma^{C\bar{C}}(e^{\pm} p)}{d{x}dQ^2} \frac{Q^4 x}{2 \pi \alpha^2 Y_{+}} = F_2^{C\bar C} - \frac{y^2}{Y_{+}}  F_L^{C\bar C}~.
    \label{eq:RNCheavy}
\end{eqnarray}
A detailed study of the inclusive deep inelastic ${e^\pm}p$ scattering cross sections, charm production reduced cross section, generalized structure functions for NC and CC deep inelastic ${e^\pm}p$ scattering and other related parameters can be found in Ref.~\citep{Vafaee:2017nze}.    

\section{\label{charm mass}Charm-quark mass in the pQCD approach}
From the theoretical point of view, the reduced charm production cross section is obtained by convolution of matrix elements with PDFs. On the other hand, PDFs are extracted from inclusive deep inelastic ${e^\pm}p$ scattering cross sections. Accordingly, both matrix elements and proton PDFs strictly depend on the $c$-quark mass~\cite{Alekhin:2012un,Alekhin:2017kpj,Kataev:2018fvx,Marquard:2016vmy,Marquard:2015qpa,Beneke:1998rk}.

The non-observation of free quarks is explained by the hypothesis of color confinement, which states that colored objects are always confined to color singlet states and that no objects with non-zero color charge can propagate as free particles~\cite{Brodsky:2017qno,Brodsky:2017tyf,Weng:2017ian,Bravina:2014jaw}. Accordingly, different definitions of the charm-quark mass $m_c$ such as the pole mass $m_c^{\rm pole}$ and \MSbar running mass $m_c(\mu_r)$ are available. In pQCD, the pole mass is defined as the mass at the position of the pole in the $c$-quark propagator and \MSbar running mass is the charm mass which is evaluated at the renormalization scale $\mu_r$~\cite{Vafaee:2019hwf}. Each definition has own advantages and disadvantages. The pole mass $m_c^{\rm pole}$ is a gauge invariant quantity and is well defined in any finite order of pQCD. But it has an intrinsic uncertainty of order $\frac{\Lambda_{\rm QCD}}{m_c}$, where $\Lambda_{QCD} \sim 0.25$ MeV~ is the QCD scale. The \MSbar running mass avoid this problem and its relation with pole mass $m_c^{\rm pole}$ is given by
\begin{equation}
	m_c^{\rm pole}=m_c(m_c)\left(1+\frac{4\alpha_s(m_c)}{3\pi}\right)~,
\label{eq:th:runpolmass}
\end{equation}  
where $m_c(m_c)$ is the \MSbar running mass evaluated at the scale $\mu_r=m_c$.

\section{\label{methodology}Data Set and Methodology}         
In this NNLO pQCD analysis, we use three different data sets: the HERA run I and II combined NC and CC deep $e^{\pm}p$ scattering cross sections (HC)~\cite{Abramowicz:2015mha}, the very recently charm production cross section H1-ZEUS combined data (HCC)~\cite{H1:2018flt} and the full five LHCb charm production cross section data sets at $\sqrt{s} = 7$~TeV~\cite{Aaij:2013mga}.

 To determine and study the pure impact of charm-quark pole mass $m_c^{\rm pole}$ on proton PDFs and fit-quality we make six different fits in two separate steps as follow:

 At the first step we fixed the charm-quark mass to $m_c = 1.257$~GeV and make three different fits with $13$ free parameters based on HCC, LHCb and HCC$+$LHCb charm production cross section data to investigate the pure impact of these data sets on the full HERA run I and II combined NC and CC deep $e^{\pm}p$ scattering data, as the central proton PDFs.
 
 At the second step we consider the charm-quark mass $m_c^{\rm pole}$ as an extra free parameter of the pQCD Lagrangian and refit the above fit procedures but this time with $14$ free parameters to determine simultaneously the charm-quark pole mass $m_c^{\rm pole}$ and pure impact of $c$-mass on the proton PDFs for the first time at the NNLO corrections. 
 
 Depending on the initial set-up of a QCD analysis, different approaches can be taken for treatment of the heavy quarks contribution~\cite{Lai:2010vv,Ball:2008by,Mironov:2009uv,Collins:1998rz,Martin:2006qz,Forte:2010ta,Martin:2009iq,Thorne:2006qt,Thorne:2012az}. 
 
 To include the charm-quark contribution, we use very recently updated Fixed Flavor number scheme from Alekhin, Blumlein and Moch (FF ABM) as implemented in the xFitter package as a powerful QCD framework~\cite{xFitter,Vafaee:2019nmo,Vafaee:2019yec,Shokouhi:2018gie,Vafaee:2018ehy,Vafaee:2017jnt,Vafaee:2016jxl}.
 
  Baed on our NNLO pQCD set-up, the FF ABM scheme provides most reliable results and best fit-quality in the phase space of HCC and LHCb charm production data. In the fixed flavor number scheme heavy quarks are considered as massive at all scales but they do not considered as partons within the proton. The number of active flavors is fixed to three for $c$-quark and is fixed to four for $b$-quark. Updated variants of FFN scheme govern both charm-quark pole mass and \MSbar running mass, however the calculations of this QCD analysis use FF ABM variant and is developed based on the charm-quark pole mass $m_c^{\rm pole}$.
 
 To parameterized the proton PDFs, we use the HERAPDF standard functional form at the initial scale of the QCD evolution $Q^2_0= 1.9$ GeV$^2$ as:
\begin{equation}
 xf(x) = A x^{B} (1-x)^{C} (1 + D x + E x^2)~~,
\label{eqn:pdf}
\end{equation}
with $13$ central free parameters and $m_c$ as another extra free parameter. A detailed review of HERAPDF standard functional form and its related parameters has been reported in Ref.~\cite{Vafaee:2017nze}. 

The initial set-up of this QCD analysis is based on the following additional parameters: The strong coupling constant is fixed to $\alpha_s^{{\rm NNLO}}(M^2_Z) = 0.118$~\cite{Vafaee:2017nze}, the strangeness suppression factor is fixed to $f_{s}=0.4$~\cite{Abramowicz:2015mha}, the initial value of charm-quark pole mass is set to $m_c^{\rm pole} = 1.257$~GeV~ and then varied in steps of 0.001~\cite{xFitter} and finally the theory type based on the DGLAP collinear evolution mode~\cite{Botje:2010ay,DGLAP}.

\section{\label{results}Results}
Table~\ref{tab:bdata}, shows the experiments with correlated ${\chi^2}$ and  ${\chi^2}_{Total}$ per degrees of freedom (dof) for each experiment corresponding to three different HCC, LHCb and HCC$+$LHCb data sets, when the charm-quark mass is fixed to $m_c = 1.257$~GeV. 

\begin{table}[h]
\begin{center}
\begin{tabular}{|l|c|c|c|c|}
\hline
\hline
{Scheme} & \multicolumn{3}{c|}{ {charm-quark mass is fixed to $m_c = 1.257$~GeV} }    \\ \hline
 {Experiment} & {$~~~~$HCC$~~~~$} & { $~~~$LHCb$~~~$} & {HCC$+$LHCb} \\ \hline
  HC CC $e^{+}p$~\cite{Abramowicz:2015mha} & 56 / 39& 63 / 39& 63 / 39 \\ 
  HC CC $e^{-}p$~\cite{Abramowicz:2015mha} & 51 / 42& 50 / 42& 50 / 42 \\ 
  HC NC $e^{-}p$~\cite{Abramowicz:2015mha} & 218 / 159& 224 / 159& 224 / 159 \\ 
  HC NC $e^{+}p$ 460~\cite{Abramowicz:2015mha} & 213 / 204& 211 / 204& 211 / 204 \\ 
  HC NC $e^{+}p$ 575~\cite{Abramowicz:2015mha} & 213 / 254& 210 / 254& 211 / 254 \\
  HC NC $e^{+}p$ 820~\cite{Abramowicz:2015mha} &  63 / 70& 61 / 70& 62 / 70 \\ 
  HC NC $e^{+}p$ 920~\cite{Abramowicz:2015mha} & 427 / 377& 425 / 377& 426 / 377 \\ \hline
 {HCC}~\cite{H1:2018flt} & 41 / 47& - & 40 / 47 \\ \hline
  {LHCb 7TeV Dzero}~\cite{Aaij:2013mga} & - & 392 / 38& 389 / 38  \\ 
  {LHCb 7TeV Dch}~\cite{Aaij:2013mga} & - & 117 / 37& 119 / 37  \\ 
  {LHCb 7TeV Dstar}~\cite{Aaij:2013mga} & - & 85 / 31& 87 / 31  \\ 
  {LHCb 7TeV Ds}~\cite{Aaij:2013mga} & - & 26 / 28& 26 / 28  \\ 
  {LHCb 7TeV Lambdac}~\cite{Aaij:2013mga} & - & 5.1 / 6& 5.2 / 6  \\ \hline   
 { Correlated ${\chi^2}$} & 157& 157& 195 \\ \hline
{${\frac{{\chi^2}_{Total}}{dof}}$} & ${\frac{1410}{1179}}$  & ${\frac{2029}{1272}}$ &  ${\frac{2078}{1319}}$  \\ \hline
\hline
    \end{tabular}
\vspace{-0.0cm}
\caption{\label{tab:bdata}{Experiments with correlated ${\chi^2}$ and  ${\chi^2}_{Total}$ per degrees of freedom (dof) for each experiment corresponding to three different HCC, LHCb and HCC$+$LHCb data sets, when the charm-quark mass is fixed to $m_c = 1.257$~GeV.}}
\vspace{-0.4cm}
\end{center}
\end{table}

 In the Table~\ref{tab:fqbdata}, we compare the pure impact of the HCC, LHCb and HCC$+$LHCb data on the fit quality, when the charm-quark mass is fixed to $m_c = 1.257$~GeV. As we can see from numerical results of Table~\ref{tab:fqbdata} the best fit quality is corresponding to HCC data.

\begin{table}[h]
\begin{center}
\begin{tabular}{|l|c|c|c|c|}
\hline
\hline
 \multicolumn{3}{|c|}{ {charm-quark mass is fixed to $m_c = 1.257$~GeV} }    \\ \hline
 {Experiment} & {$\frac{{\chi^2}_{Total}}{dof}$} & {fit quality} \\ \hline 
  {HCC} & {${1410/1179}$}& $1.19$  \\
  {LHCb} & {${2029/1272}$}& $1.59$  \\
  {HCC$+$LHCb} & {${2078/1319}$}& $1.57$  \\ \hline
    \end{tabular}
\vspace{-0.0cm}
\caption{\label{tab:fqbdata}{ Comparison the pure impact of the HCC, LHCb and HCC$+$LHCb data on the fit quality, when the charm-quark mass is fixed to $m_c = 1.257$~GeV.}}
\vspace{-0.4cm}
\end{center}
\end{table}

 In Table~\ref{tab:bpar}, we present NNLO numerical values of $13$ free central parameters and their uncertainties for  the $xu_v$, $xd_v$, sea and gluon distributions at the input scale of $Q^2_0 = 1.9$~GeV$^2$ for three different HCC, LHCb and HCC$+$LHCb data sets.

\begin{table}[h]
\begin{center}
\begin{tabular}{|l|c|c|c|c|}
\hline
\hline
\multicolumn{4}{|c|}{ {charm-quark mass is fixed to $m_c = 1.257$~GeV} }    \\ \hline
 {Parameter} & {$~~~~$HCC$~~~~$} & { $~~~$LHCb$~~~$} & {HCC$+$LHCb} \\ \hline
  ${B_{u_v}}$ & $0.865 \pm 0.033$& $0.831 \pm 0.024$& $0.816 \pm 0.025$  \\ 
  ${C_{u_v}}$ & $4.392 \pm 0.076$& $4.411 \pm 0.087$& $4.415 \pm 0.083$  \\ 
  $E_{u_v}$ & $9.4 \pm 1.3$& $9.6 \pm 1.2$& $10.2 \pm 1.3$  \\ \hline
  ${B_{d_v}}$ & $1.056 \pm 0.093$& $0.978 \pm 0.078$& $0.961 \pm 0.079$  \\ 
  $C_{d_v}$ & $4.47 \pm 0.37$& $4.67 \pm 0.39$& $4.58 \pm 0.38$ \\ \hline
  $C_{\bar{U}}$ & $3.77 \pm 0.54$& $2.35 \pm 0.31$& $2.45 \pm 0.32$  \\ 
  $A_{\bar{D}}$ & $0.1801 \pm 0.0091$& $0.1629 \pm 0.0072$& $0.1682 \pm 0.0075$  \\  
  $B_{\bar{D}}$ & $-0.1691 \pm 0.0063$& $-0.1807 \pm 0.0056$& $-0.1757 \pm 0.0056$  \\ 
  $C_{\bar{D}}$ & $5.7 \pm 1.0$& $5.69 \pm 0.87$& $5.80 \pm 0.90$ \\ \hline
  $B_g$ & $0.23 \pm 0.13$& $-0.196 \pm 0.026$& $-0.192 \pm 0.027$  \\ 
  $C_g$ & $4.89 \pm 0.77$& $3.23 \pm 0.32$& $2.99 \pm 0.31$  \\ 
  $A_g'$ &  $2.69 \pm 0.51$& $1.75 \pm 0.20$& $1.64 \pm 0.19$  \\ 
  ${B_g'}$ & $0.101 \pm 0.058$& $-0.161 \pm 0.026$& $-0.157 \pm 0.028$  \\ 
  \hline 
  {$\alpha_s^{{\rm NNLO}}(M^2_Z)$} & $ 0.118$ & $ 0.118$ & $ 0.118$ \\ \hline
  {$m_c$} & ${ 1.257 }$ & $ { 1.257 }$ & ${ 1.257 }$  \\  \hline
\hline
    \end{tabular}
\vspace{-0.0cm}
\caption{\label{tab:bpar}{ {The NNLO numerical values of $13$ free central parameters and their uncertainties for  the $xu_v$, $xd_v$, sea and gluon distributions at the input scale of $Q^2_0 = 1.9$~GeV$^2$ for three different HCC, LHCb and HCC$+$LHCb data sets.}}}
\vspace{-0.4cm}
\end{center}
\end{table}

 According to the numerical values of Table~\ref{tab:bpar} and three different fit qualities from Table~\ref{tab:fqbdata}, we expect to see dramatically impact of the HCC, LHCb and HCC$+$LHCb data on the shape of the gluon distribution and some of its ratios.

\begin{figure*}
\includegraphics[width=0.49\textwidth]{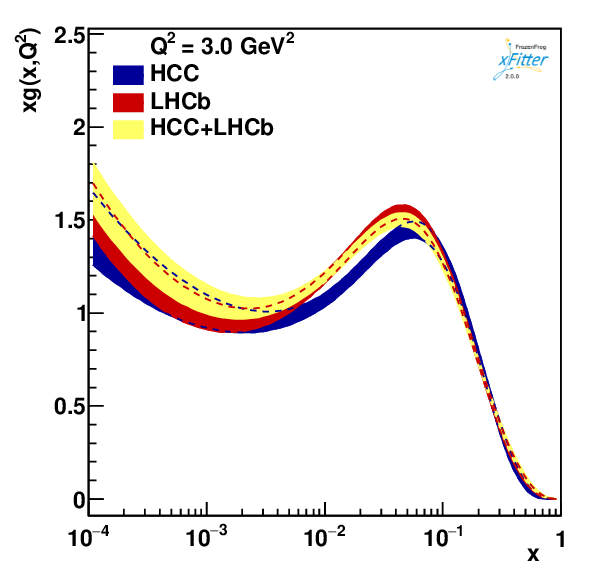}
\includegraphics[width=0.49\textwidth]{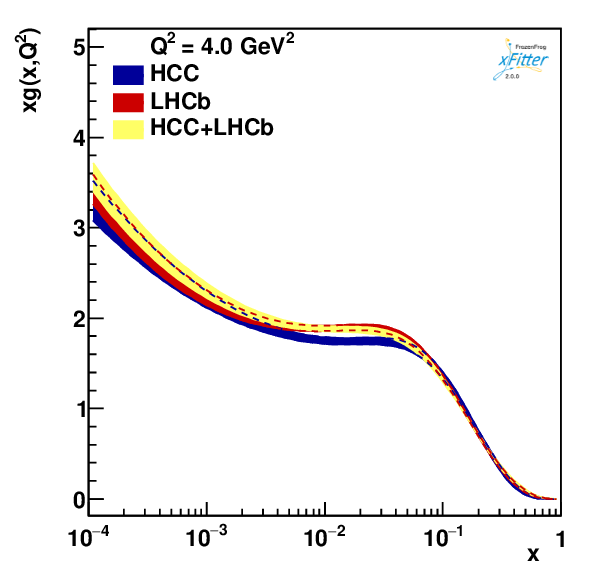}

\includegraphics[width=0.49\textwidth]{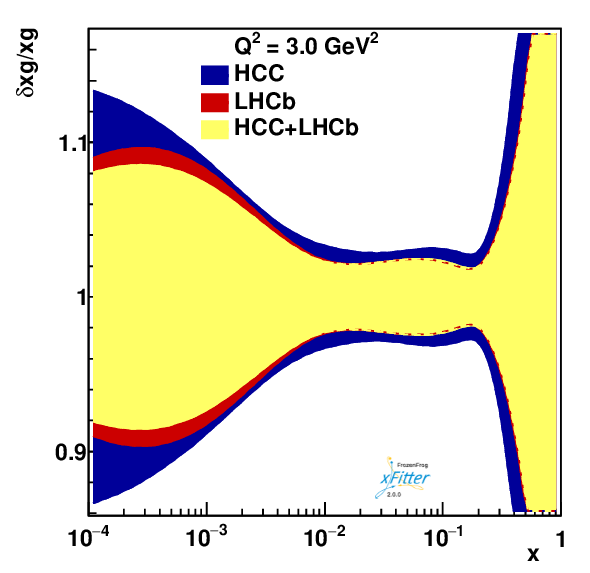}
\includegraphics[width=0.49\textwidth]{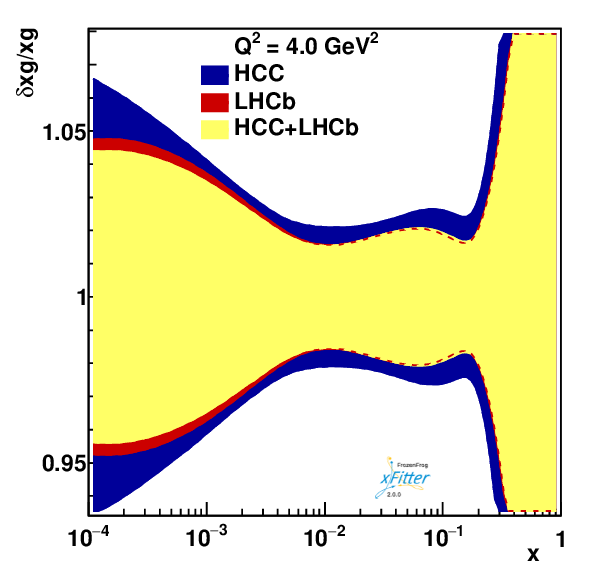}

\includegraphics[width=0.49\textwidth]{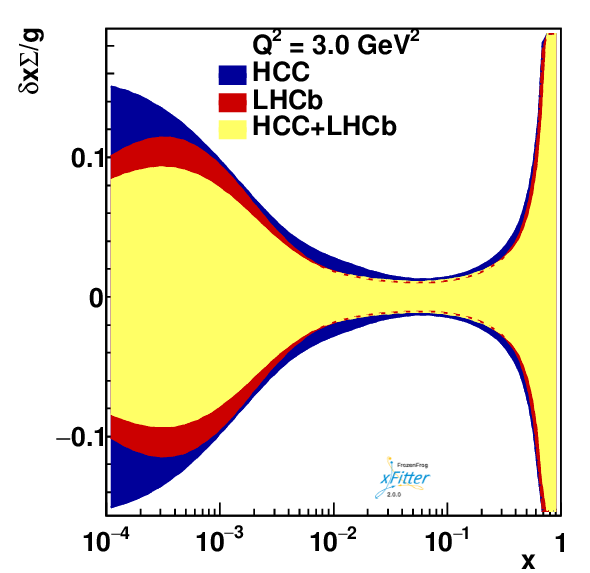}
\includegraphics[width=0.49\textwidth]{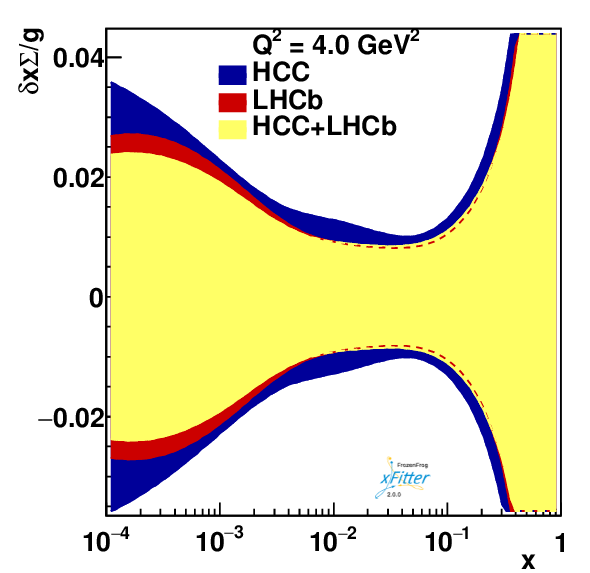}

\caption{The gluon PDFs (two upper), the partial of gluon PDFs (two middle) and the partial ratio of $\Sigma$-PDFs over gluon distributions (two lower) as extracted from three different HCC (blue), LHCb (red) and HCC$+$LHCb (yellow) data sets.}
\label{fig:1}
\end{figure*}

 In Fig.~\ref{fig:1}, we show the gluon PDFs (two upper), the partial of gluon PDFs (two middle) and the partial ratio of $\Sigma$-PDFs over gluon distributions  (two lower) as extracted from three different HCC (blue), LHCb (red) and HCC$+$LHCb (yellow) data sets.

 As can be seen from Fig.~\ref{fig:1} the best improvement of uncertainty error bands is corresponding to HCC$+$LHCb data with yellow color, which in turn strictly confirms that the DIS measurements depend on the various phenomenological input data and knowledge of the PDFs. Also, the pure impact of LHCb charm production data in improvement of the gluon distribution and some of its ratios (red color) is better than the pure impact of HCC data (blue color).

 Now in the second step we consider the charm-quark mass as an extra free parameter and repeat our previous fit procedures but this time with $14$ free parameters to determine both charm-quark mass and pure impact of $c$-mass on the shape of the proton PDFs and fit quality.

Table~\ref{tab:mdata}, shows the experiments with correlated ${\chi^2}$ and  $\frac{{\chi^2}_{Total}}{dof}$ for each experiment corresponding to three different HCC, LHCb and HCC$+$LHCb data sets, when the charm-quark mass $m_c$ is taken as an extra pQCD free parameter.

\begin{table}[h]
\begin{center}
\begin{tabular}{|l|c|c|c|c|}
\hline
\hline
{Scheme} & \multicolumn{3}{c|}{ {$m_c$ is taken as an extra pQCD free parameter} }    \\ \hline
 {Experiment} & {$~~~~$HCC$~~~~$} & { $~~~$LHCb$~~~$} & {HCC$+$LHCb} \\ \hline
  HC CC $e^{+}p$~\cite{Abramowicz:2015mha} & 55 / 39& 57 / 39& 58 / 39 \\ 
  HC CC $e^{-}p$~\cite{Abramowicz:2015mha} & 51 / 42& 51 / 42& 51 / 42 \\ 
  HC NC $e^{-}p$~\cite{Abramowicz:2015mha} & 218 / 159& 223 / 159& 225 / 159 \\ 
  HC NC $e^{+}p$ 460~\cite{Abramowicz:2015mha} & 213 / 204& 209 / 204& 208 / 204 \\ 
  HC NC $e^{+}p$ 575~\cite{Abramowicz:2015mha} & 213 / 254& 211 / 254& 210 / 254 \\
  HC NC $e^{+}p$ 820~\cite{Abramowicz:2015mha} &  63 / 70& 63 / 70& 63 / 70 \\ 
  HC NC $e^{+}p$ 920~\cite{Abramowicz:2015mha} & 425 / 377& 418 / 377& 420 / 377 \\ \hline
 {HCC}~\cite{H1:2018flt} & 43 / 47& - & 58 / 47 \\ \hline
  {LHCb 7TeV Dzero}~\cite{Aaij:2013mga} & - & 108 / 38& 128 / 38  \\ 
  {LHCb 7TeV Dch}~\cite{Aaij:2013mga} & - & 69 / 37& 73 / 37  \\ 
  {LHCb 7TeV Dstar}~\cite{Aaij:2013mga} & - & 50 / 31& 54 / 31  \\ 
  {LHCb 7TeV Ds}~\cite{Aaij:2013mga} & - & 30 / 28& 28 / 28  \\ 
  {LHCb 7TeV Lambdac}~\cite{Aaij:2013mga} & - & 7.0 / 6& 6.3 / 6  \\ \hline   
 { Correlated ${\chi^2}$} & 157& 215& 261 \\ \hline
{${\frac{{\chi^2}_{Total}}{dof}}$} & ${\frac{1438}{1178}}$  & ${\frac{1710}{1271}}$ &  ${\frac{1844}{1318}}$  \\ \hline
\hline
    \end{tabular}
\vspace{-0.0cm}
\caption{\label{tab:mdata}{Experiments with correlated ${\chi^2}$ and  $\frac{{\chi^2}_{Total}}{dof}$ for each experiment corresponding to three different HCC, LHCb and HCC$+$LHCb data sets, when the charm-quark mass $m_c$ is taken as an extra pQCD free parameter.}}
\vspace{-0.4cm}
\end{center}
\end{table}

 In Table~\ref{tab:fqmdata} we compare the pure impact of the HCC, LHCb and HCC$+$LHCb data sets on the fit-quality, when the charm-quark mass $m_c$ is taken as an extra pQCD free parameter. As we can see from the numerical results of Table~\ref{tab:fqmdata} the best fit quality is corresponding to the HCC data set.

\begin{table}[h]
\begin{center}
\begin{tabular}{|l|c|c|c|c|}
\hline
\hline
 \multicolumn{3}{|c|}{ {charm-quark mass is taken as an extra free parameter}}    \\ \hline
 {Experiment} & {$\frac{{\chi^2}_{Total}}{dof}$} & {fit quality} \\ \hline 
  {HCC} & {${ 1438/1178}$}& $1.22$  \\
  {LHCb} & {${1710/1271}$}& $1.34$  \\
  {HCC$+$LHCb} & {${1844/1318}$}& $1.40$  \\ \hline
    \end{tabular}
\vspace{-0.0cm}
\caption{\label{tab:fqmdata}{Comparison the pure impact of the HCC, LHCb and HCC$+$LHCb data on the fit quality, when the charm-quark mass $m_c$ is taken as an extra pQCD free parameter.}}
\vspace{-0.4cm}
\end{center}
\end{table}

\begin{figure*}
\includegraphics[width=0.49\textwidth]{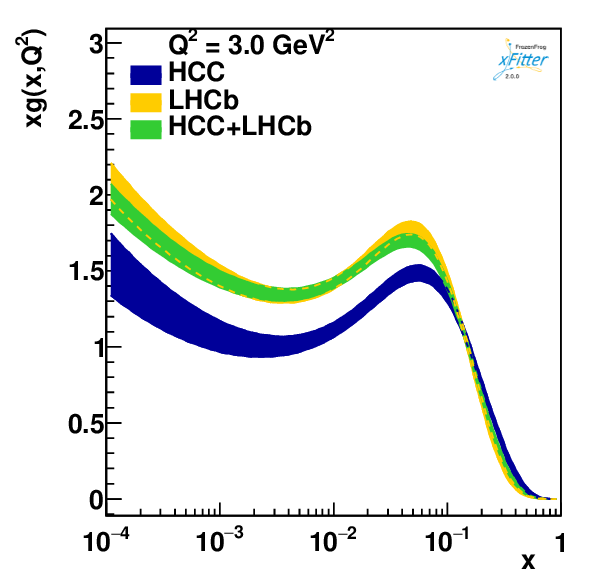}
\includegraphics[width=0.49\textwidth]{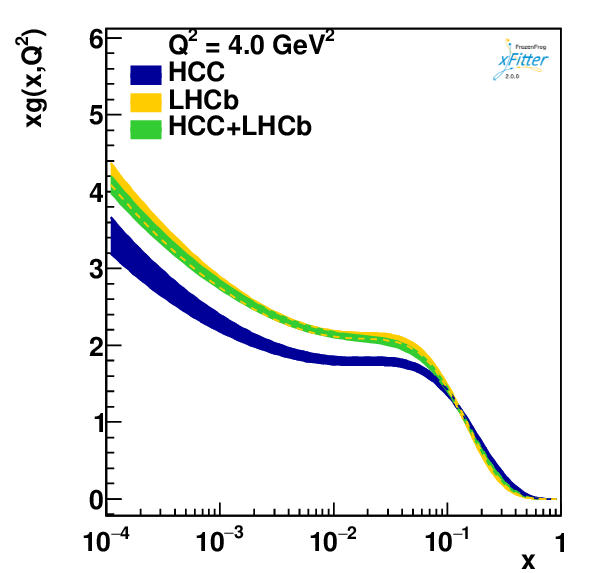}

\includegraphics[width=0.49\textwidth]{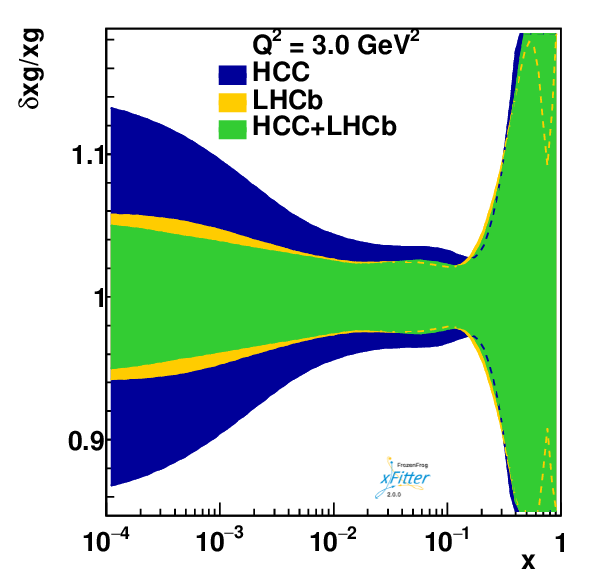}
\includegraphics[width=0.49\textwidth]{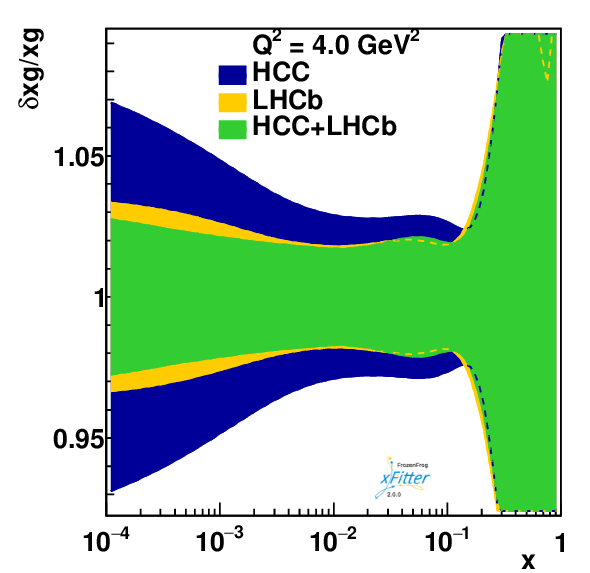}

\includegraphics[width=0.49\textwidth]{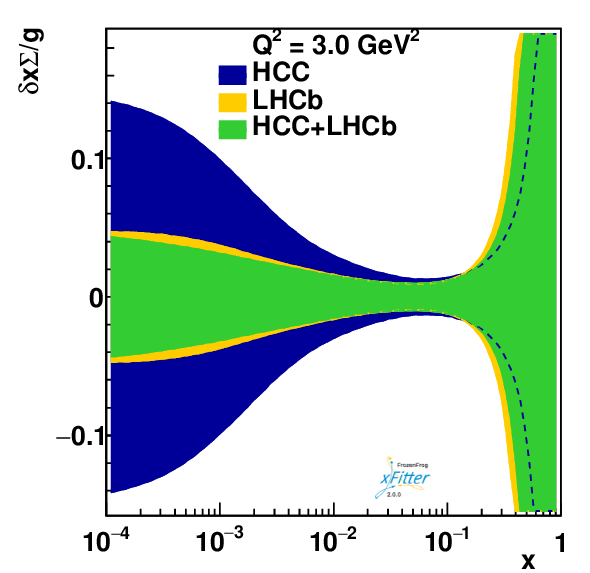}
\includegraphics[width=0.49\textwidth]{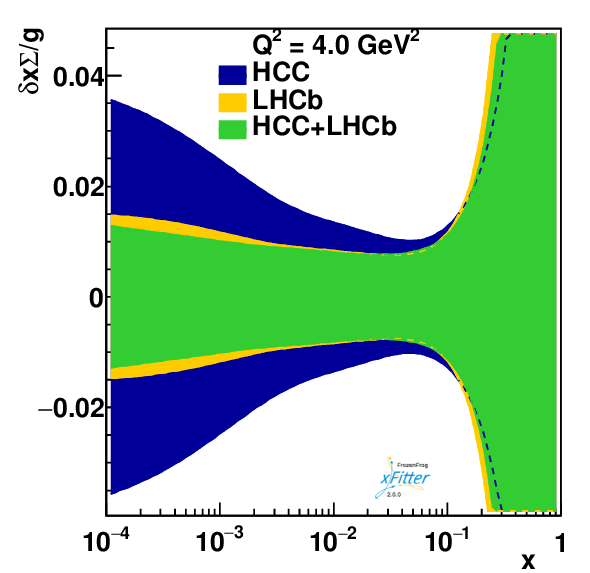}
\caption{The gluon PDFs (two upper), the partial of gluon PDFs (two middle) and the partial ratio of $\Sigma$-PDFs over gluon distributions  (two lower) as extracted from three different HCC (blue), LHCb (red) and HCC$+$LHCb (yellow) data sets.}
\label{fig:2}
\end{figure*}

In Fig.~\ref{fig:2}, we show the gluon PDFs (two upper), the partial of gluon PDFs (two middle) and the partial ratio of $\Sigma$-PDFs over gluon distributions  (two lower) as extracted from three different HCC (blue), LHCb (red) and HCC$+$LHCb (yellow) data sets.

 It is clear from the Fig.~\ref{fig:2} that the best improvement of uncertainty error bands is corresponding to HCC$+$LHCb data with green color. Also, the pure impact of LHCb charm production data in improvement of the gluon distribution and some of its ratios (yellow color) is significantly better than the pure impact of HCC data (blue color).
 
 As one can find from the Fig.~\ref{fig:2} (top $2$ plots), the blue lines (HCC) clearly are separated from the yellow lines (LHCb). This issue physically means that, the actual charm-quark pole mass $m_c^{\rm pole}$ increases at higher energies in the scattering process.
 
 According to the absolute relative change of $\chi^2$ function which is defined by $\vert\frac{\chi^2_{\rm final}-\chi^2_{\rm initial}}{\chi^2_{\rm initial}}\vert$~, we may conclude from numerical results of the Tables~\ref{tab:fqbdata} and \ref{tab:fqmdata}:
 
\begin{itemize}
\item Relative improvement in the quality of the fit corresponding to HCC data is $\vert{\frac{1.22-1.19}{1.19}}\vert\sim 2.5$~\%, without and with the charm-quark mass $m_c$ is taken as an extra pQCD free parameter. 

\item Relative improvement in the quality of the fit corresponding to LHCb data is $\vert{\frac{1.34-1.59}{1.59}}\vert\sim 15.7$~\%, with and without the charm-quark mass $m_c$ is taken as an extra pQCD free parameter.

\item Relative improvement in the quality of the fit corresponding to HCC$+$LHCb data is $\vert{\frac{1.40-1.57}{1.57}}\vert\sim 10.8$~\%, with and without the charm-quark mass $m_c$ is taken as an extra pQCD free parameter.
\end{itemize} 
 
 Since the $\chi^2$ function is a measure of the agreement between data and theory models, we led to this fact that: deep inelastic ${e^\pm}p$ scattering measurements depend on the various phenomenological input data. On the other hand, dramatically improvement of the error bands of the gluon content of the proton corresponding to LHCb data, strictly confirms these results.

 Determination of the $14$ free fit parameters, including $13$ central proton PDF parameters and charm-quark mass $m_c$ as an extra pQCD free parameter are presented in Table~\ref{tab:mpar}.

\begin{table}[h]
\begin{center}
\begin{tabular}{|l|c|c|c|c|}
\hline
\hline
\multicolumn{4}{|c|}{ {charm-quark mass is taken as an extra free parameter} }    \\ \hline
 {Parameter} & {$~~~~$HCC$~~~~$} & { $~~~$LHCb$~~~$} & {HCC$+$LHCb} \\ \hline
  ${B_{u_v}}$ & $0.865 \pm 0.032$& $0.860 \pm 0.026$& $0.834 \pm 0.024$  \\ 
  ${C_{u_v}}$ & $4.387 \pm 0.079$& $4.39 \pm 0.11$& $4.45 \pm 0.10$  \\ 
  $E_{u_v}$ & $9.3 \pm 1.3$& $8.4 \pm 1.3$& $9.6 \pm 1.3$  \\ \hline
  ${B_{d_v}}$ & $1.050 \pm 0.091$& $1.012 \pm 0.082$& $0.977 \pm 0.079$  \\ 
  $C_{d_v}$ & $4.47 \pm 0.37$& $4.79 \pm 0.39$& $4.72 \pm 0.39$ \\ \hline
  $C_{\bar{U}}$ & $3.59 \pm 0.56$& $2.35 \pm 0.30$& $2.21 \pm 0.28$  \\ 
  $A_{\bar{D}}$ & $0.1834 \pm 0.0096$& $0.1936 \pm 0.0078$& $0.1897 \pm 0.0074$  \\  
  $B_{\bar{D}}$ & $-0.1669 \pm 0.0066$& $-0.1589 \pm 0.0049$& $-0.1601 \pm 0.0046$  \\ 
  $C_{\bar{D}}$ & $5.7 \pm 1.0$& $5.36 \pm 0.86$& $5.49 \pm 0.89$ \\ \hline
  $B_g$ & $0.24 \pm 0.14$& $0.212 \pm 0.096$& $0.151 \pm 0.090$  \\ 
  $C_g$ & $5.27 \pm 0.87$& $8.07 \pm 0.62$& $6.57 \pm 0.62$  \\ 
  $A_g'$ &  $2.95 \pm 0.60$& $5.77 \pm 0.86$& $3.77 \pm 0.60$  \\ 
  ${B_g'}$ & $0.112 \pm 0.063$& $0.162 \pm 0.057$& $0.097 \pm 0.052$  \\ 
  \hline 
  {$\alpha_s^{{\rm NNLO}}(M^2_Z)$} & $ 0.118$ & $ 0.118$ & $ 0.118$ \\ \hline
  {$m_c$} & ${1.331 \pm 0.058}$ & $ { 1.760 \pm 0.028 }$ & ${ 1.655 \pm 0.022 }$  \\  \hline
\hline
    \end{tabular}
\vspace{-0.0cm}
\caption{\label{tab:mpar}{ {The NNLO numerical values of $14$ free fit parameters, including $13$ central proton PDF parameters and charm-quark mass $m_c$ as an extra pQCD free parameter for three different HCC, LHCb and HCC$+$LHCb data sets.}}}
\vspace{-0.4cm}
\end{center}
\end{table}

 As we expected, the best uncertainty improvement from the central value of $c$-quark mass is $m_c = 1.655 \pm 0.022$, corresponding to HCC$+$LHCb  data sets.

 The comparison of these results with the measurements
from the PDG world average~\cite{Agashe:2014kda} shows a very good agreement with the expected charm-quark mass.

\begin{figure*}
\includegraphics[width=0.49\textwidth]{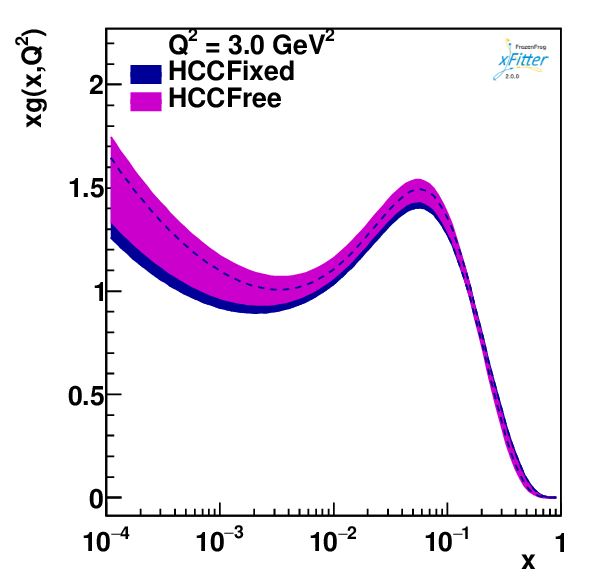}
\includegraphics[width=0.49\textwidth]{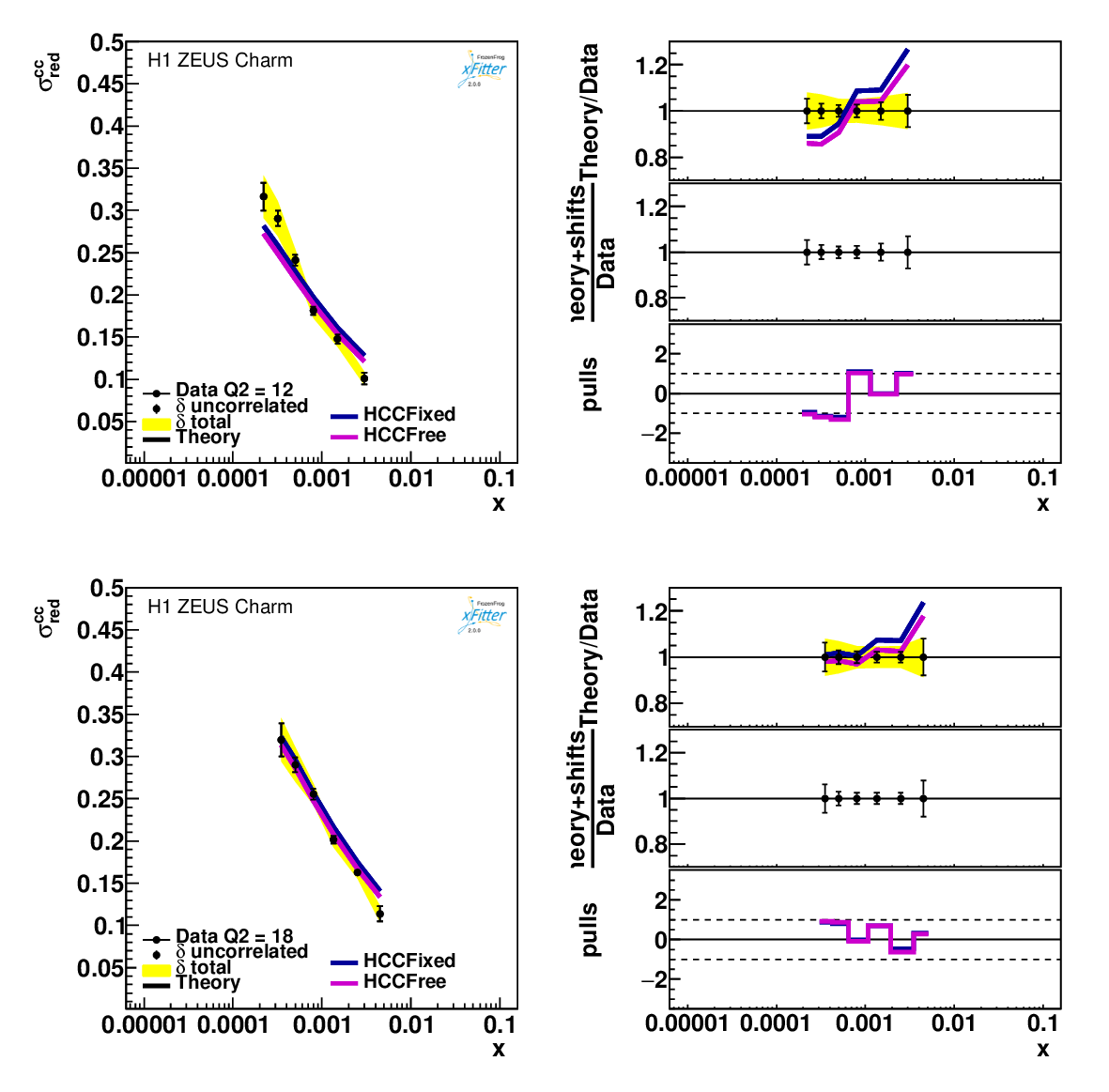}

\includegraphics[width=0.49\textwidth]{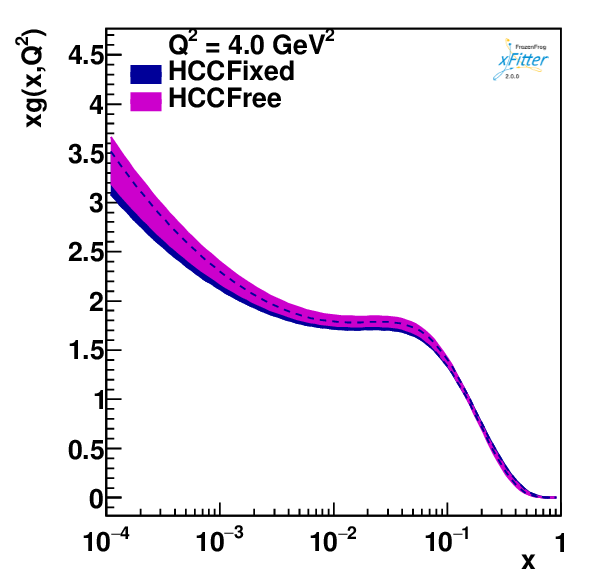}
\includegraphics[width=0.49\textwidth]{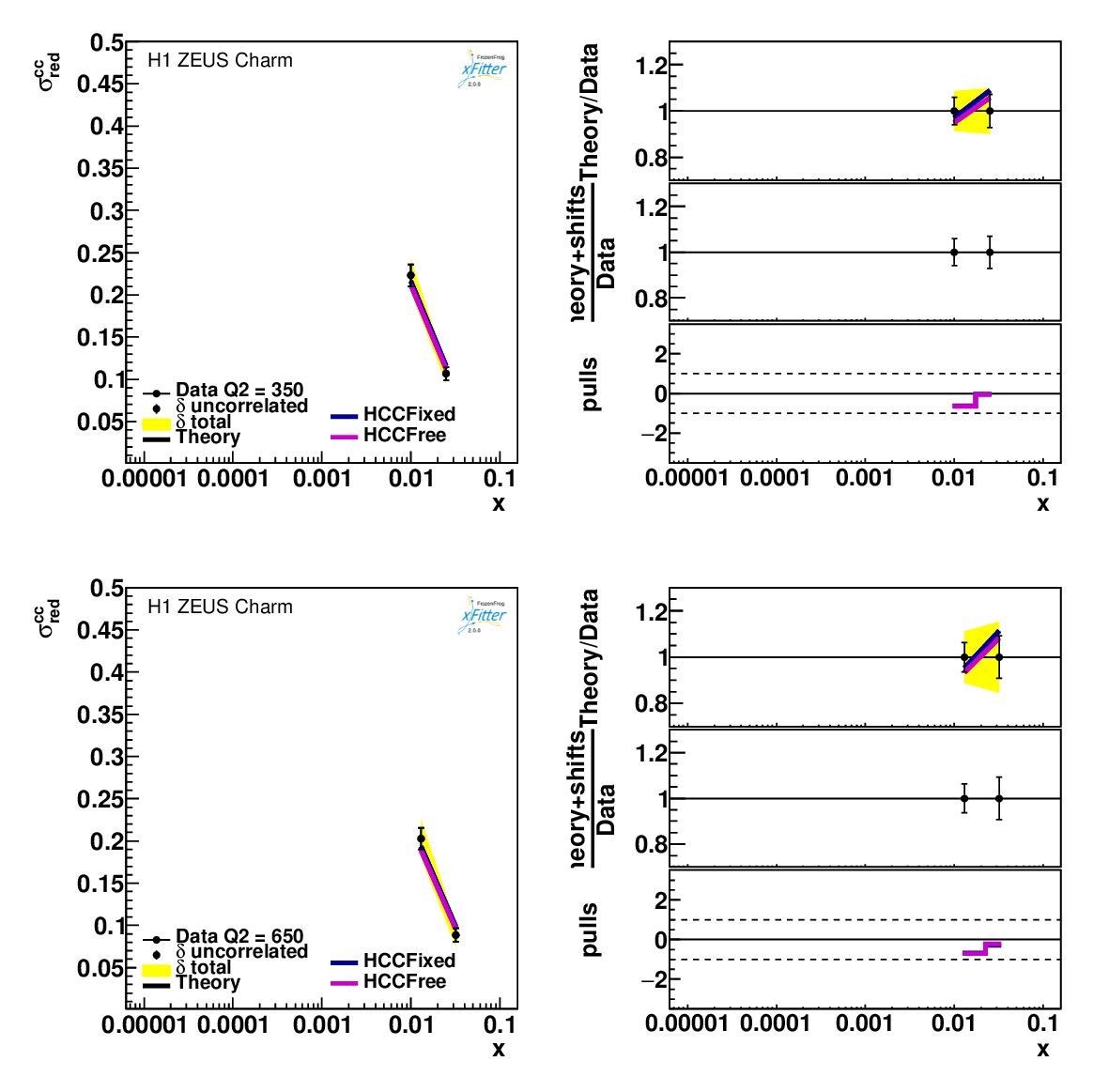}

\caption{The pure impact of $c$-mass on gluon distribution and consistency between pQCD theory predictions and the phenomenology of experimental data in determination of the charm-quark pole mass $m_c^{\rm pole}$ at the NNLO corrections in three separate panels, include of pulls, $\frac{{\text {Theory}} + {\text {Shifts}}}{\text {Data}}$ and $\frac{\text {Theory}}{\text {Data}}$ corresponding to HCC data set.}
\label{fig:3}
\end{figure*}

\begin{figure*}
\includegraphics[width=0.49\textwidth]{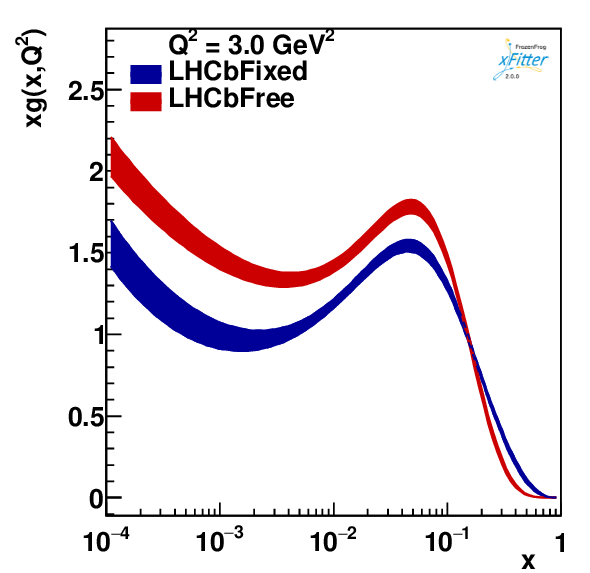}
\includegraphics[width=0.49\textwidth]{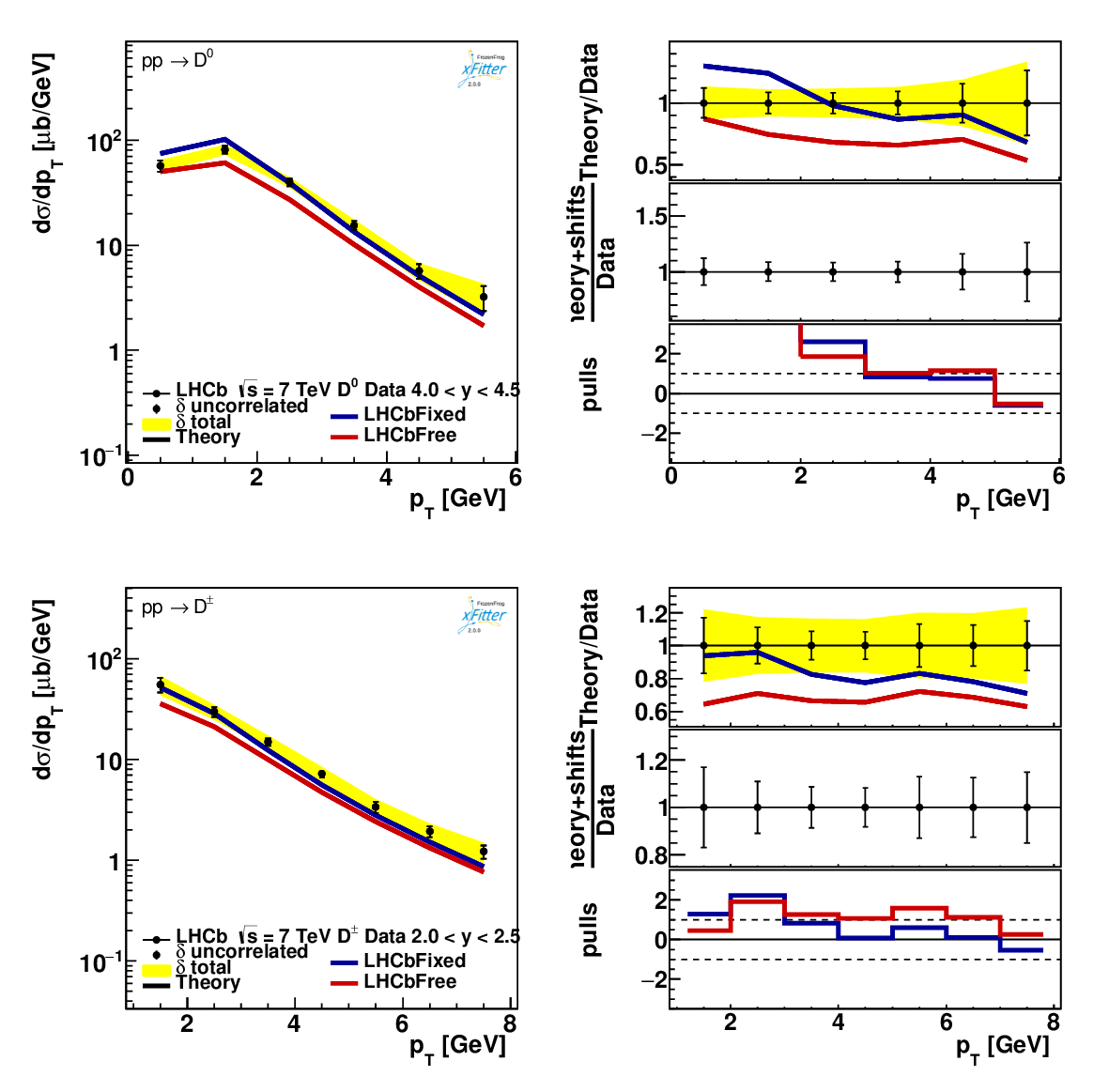}

\includegraphics[width=0.49\textwidth]{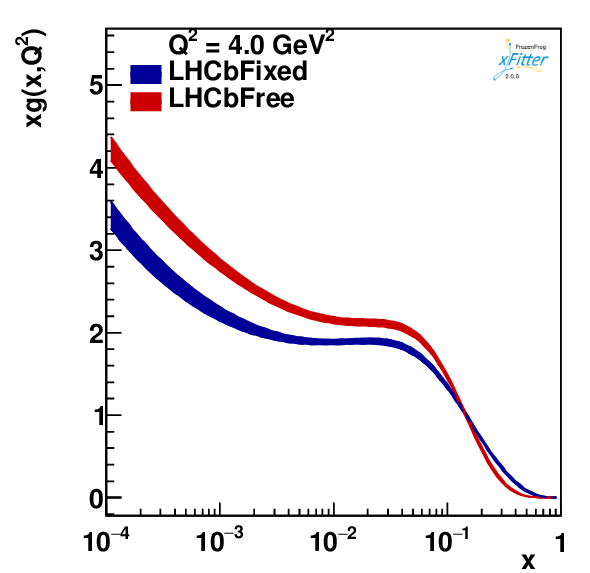}
\includegraphics[width=0.49\textwidth]{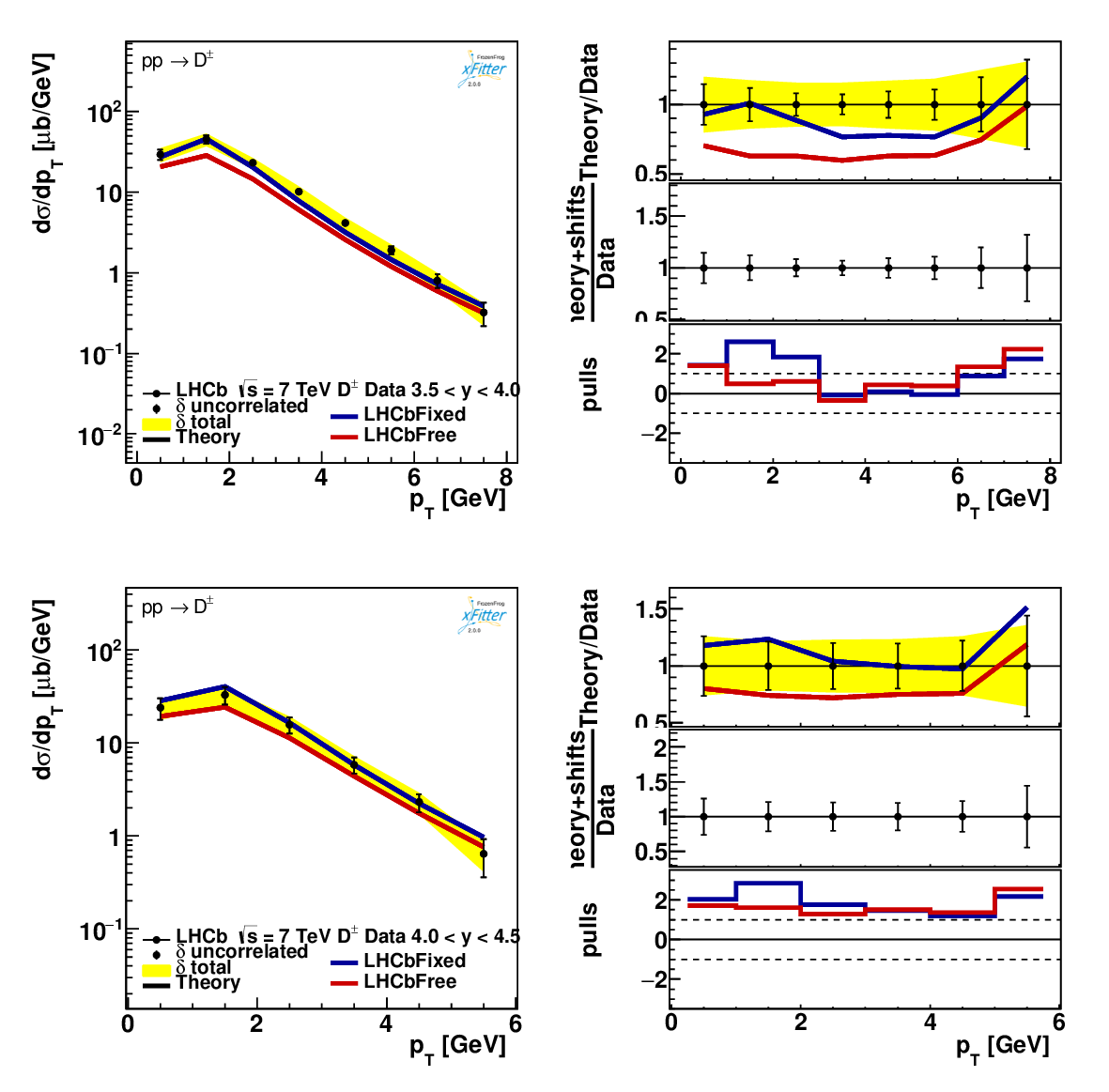}

\caption{The pure impact of $c$-mass on gluon distribution and consistency between pQCD theory predictions and the phenomenology of experimental data in determination of the charm-quark pole mass $m_c^{\rm pole}$ at the NNLO corrections in three separate panels, include of pulls, $\frac{{\text {Theory}} + {\text {Shifts}}}{\text {Data}}$ and $\frac{\text {Theory}}{\text {Data}}$ corresponding to LHCb data set.}
\label{fig:4}
\end{figure*}

 The pure impact of $c$-mass on gluon distribution and consistency between pQCD theory predictions and the phenomenology of experimental data in determination of the charm-quark pole mass $m_c^{\rm pole}$ at the NNLO corrections in three separate panels, include of pulls, $\frac{{\text {Theory}} + {\text {Shifts}}}{\text {Data}}$ and $\frac{\text {Theory}}{\text {Data}}$ corresponding to  HCC and LHCb data sets are shown in Figs.~\ref{fig:3} and~\ref{fig:4}.

\clearpage

\section{\label{Summary}Summary}
Using three deferent HCC, LHCb and HCC$+$LHCb data sets, we have simultaneously determined proton PDFs and charm-quark pole mass $m_c^{\rm pole}$ at the NNLO corrections.

 We have studied the pure impact of the three deferent HCC, LHCb and HCC$+$LHCb data sets and also pure contribution of charm-quark mass $m_c$ on the uncertainty bands of proton PDFs and fit-quality in two separate steps with following results:

\begin{itemize}

\item The best improvement of uncertainty error bands is corresponding to HCC$+$LHCb data, which strictly confirms that the deep inelastic ${e^\pm}p$ scattering measurements depend on the various phenomenological input data and knowledge of the proton PDFs.

\item Because of correlation between proton PDFs and charm-quark pole mass $m_c^{\rm pole}$ as an extra free parameter of the pQCD Lagrangian, the gluon content of the proton is dramatically sensitive to $c$-mass.

\item We have shown that, the actual charm-quark pole mass $m_c^{\rm pole}$ increases at higher energies in the deep inelastic ${e^\pm}p$ scattering measurements scattering process.

\item The best relative improvement in the quality of the fit is corresponding to LHCb data up to $15.7$~\%, with and without the charm-quark mass $m_c$ is taken as an extra pQCD free parameter.

\item The best uncertainty improvement from the central value of $c$-quark mass is $m_c = 1.655 \pm 0.022$, corresponding to HCC$+$LHCb  data sets, which emphasizes once again, the DIS measurements depend strictly on the various phenomenological input data sets.

\end{itemize} 

 In this NNLO  pQCD analysis, we presented the central role of charm-quark pole mass $m_c^{\rm pole}$ in the improvement of uncertainty band of gluon distribution and QCD fit-quality, when it is considered as an extra free parameter of the pQCD Lagrangian.

 Standard LHAPDF library files of all fit processes are available and can be obtained via e-mail from the authors.

\section{Acknowledgments}
We gratefully acknowledge Prof. V. Radescu for guidance and useful discussions about PDFs and xFitter. We are grateful to Prof. M. Botje for providing the QCDNUM package as a very fast QCD evolution program. We are also grateful to Prof. F. Olness for developing valuable heavy-flavor schemes as implemented in the xFitter package. We would like to thanks Dr. Francesco Giuli, Dr. Ivan Novikov, Dr. Oleksandr Zenaiev and Dr. Sasha Glazov from xFitter developer group for guidance and  technical support. This work is related to the ``Special Support Program for the Promotion of Scientific Authority'' in Ferdowsi University of Mashhad.



\clearpage
\begin{thebibliography}{99}
  
\bibitem{H1:2018flt}
H.~Abramowicz \textit{et al.} [H1 and ZEUS],
Eur. Phys. J. C \textbf{78}, no.6, 473 (2018)
[arXiv:1804.01019 [hep-ex]].

  
\bibitem{Aaij:2013mga} 
  R.~Aaij {\it et al.} [LHCb Collaboration],
  Nucl.\ Phys.\ B {\bf 871}, 1 (2013)
  [arXiv:1302.2864 [hep-ex]].
  
\bibitem{Alekhin:2012vu} 
  S.~Alekhin, J.~Blümlein, K.~Daum, K.~Lipka and S.~Moch,
  Phys.\ Lett.\ B {\bf 720}, 172 (2013)
  [arXiv:1212.2355 [hep-ph]].
  
\bibitem{Gao:2013wwa} 
  J.~Gao, M.~Guzzi and P.~M.~Nadolsky,
  Eur.\ Phys.\ J.\ C {\bf 73}, no. 8, 2541 (2013)
  [arXiv:1304.3494 [hep-ph]].
  
\bibitem{Tung:2006tb} 
  W.~K.~Tung, H.~L.~Lai, A.~Belyaev, J.~Pumplin, D.~Stump and C.-P.~Yuan,
  JHEP {\bf 0702}, 053 (2007)
  [hep-ph/0611254].
  
\bibitem{Aaron:2009aa} 
  F.~D.~Aaron {\it et al.} [H1 and ZEUS Collaborations],
  JHEP {\bf 1001}, 109 (2010)
  [arXiv:0911.0884 [hep-ex]].
  
\bibitem{Blumlein:2012bf} 
  J.~Blumlein,
  Prog.\ Part.\ Nucl.\ Phys.\  {\bf 69}, 28 (2013)
  [arXiv:1208.6087 [hep-ph]].
  
\bibitem{Zenaiev:2016kfl} 
  O.~Zenaiev,
  Eur.\ Phys.\ J.\ C {\bf 77}, no. 3, 151 (2017)
  [arXiv:1612.02371 [hep-ex]].
  
\bibitem{Zenaiev:2015rfa} 
  O.~Zenaiev {\it et al.} [PROSA Collaboration],
  Eur.\ Phys.\ J.\ C {\bf 75}, no. 8, 396 (2015)
  [arXiv:1503.04581 [hep-ph]].
  
\bibitem{Aaij:2013noa} 
  R.~Aaij {\it et al.} [LHCb Collaboration],
  JHEP {\bf 1308}, 117 (2013)
  [arXiv:1306.3663 [hep-ex]].
  
\bibitem{Aaij:2015bpa} 
  R.~Aaij {\it et al.} [LHCb Collaboration],
  JHEP {\bf 1603}, 159 (2016)
  Erratum: [JHEP {\bf 1609}, 013 (2016)]
  Erratum: [JHEP {\bf 1705}, 074 (2017)]
  [arXiv:1510.01707 [hep-ex]].
  
\bibitem{Aaij:2016jht} 
  R.~Aaij {\it et al.} [LHCb Collaboration],
  JHEP {\bf 1706}, 147 (2017)
  [arXiv:1610.02230 [hep-ex]].
  
\bibitem{Alves:2008zz} 
  A.~A.~Alves, Jr. {\it et al.} [LHCb Collaboration],
  JINST {\bf 3}, S08005 (2008).
  
\bibitem{Adinolfi:2012qfa} 
  M.~Adinolfi {\it et al.} [LHCb RICH Group],
  Eur.\ Phys.\ J.\ C {\bf 73}, 2431 (2013)
  [arXiv:1211.6759 [physics.ins-det]].
  
\bibitem{Vafaee:2017nze} 
  A.~Vafaee and A.~N.~Khorramian,
  Nucl.\ Phys.\ B {\bf 921}, 472 (2017)
  [arXiv:1709.08346 [hep-ph]].

  
\bibitem{Alekhin:2012un} 
  S.~Alekhin, K.~Daum, K.~Lipka and S.~Moch,
  Phys.\ Lett.\ B {\bf 718}, 550 (2012)
  [arXiv:1209.0436 [hep-ph]].
  
\bibitem{Alekhin:2017kpj} 
  S.~Alekhin, J.~Blümlein, S.~Moch and R.~Placakyte,
  Phys.\ Rev.\ D {\bf 96}, no. 1, 014011 (2017)
  [arXiv:1701.05838 [hep-ph]].
  
\bibitem{Kataev:2018fvx} 
  A.~L.~Kataev and V.~S.~Molokoedov,
  arXiv:1809.04395 [hep-ph].
  
\bibitem{Marquard:2016vmy} 
  P.~Marquard, A.~V.~Smirnov, V.~A.~Smirnov and M.~Steinhauser,
  PoS RADCOR {\bf 2015}, 094 (2016)
  [arXiv:1601.03748 [hep-ph]].
  
\bibitem{Marquard:2015qpa} 
  P.~Marquard, A.~V.~Smirnov, V.~A.~Smirnov and M.~Steinhauser,
  Phys.\ Rev.\ Lett.\  {\bf 114}, no. 14, 142002 (2015)
  [arXiv:1502.01030 [hep-ph]].
  
\bibitem{Beneke:1998rk} 
  M.~Beneke,
  Phys.\ Lett.\ B {\bf 434}, 115 (1998)
  [hep-ph/9804241].
  
\bibitem{Brodsky:2017qno} 
  S.~J.~Brodsky,
  arXiv:1709.01191 [hep-ph].
  
\bibitem{Brodsky:2017tyf}
  S.~J.~Brodsky,
  Russ.\ Phys.\ J.\  {\bf 60} (2017) no.3,  399.
  
\bibitem{Weng:2017ian} 
  Z.~H.~Weng,
  Adv.\ Math.\ Phys.\  {\bf 2017}, 9876464 (2017)
  [arXiv:1704.02240 [physics.gen-ph]].
  
\bibitem{Bravina:2014jaw} 
  L.~Bravina, A.~Di Giacomo, Y.~Foka and S.~Kabana,
  EPJ Web Conf.\  {\bf 70}, 00019 (2014).
  
\bibitem{Vafaee:2019hwf}
A.~Vafaee and K.~Javidan,
Mod. Phys. Lett. A \textbf{35}, no.30, 2050253 (2020)
[arXiv:1909.00796 [hep-ph]].
  
\bibitem{Abramowicz:2015mha} 
  H.~Abramowicz {\it et al.} [H1 and ZEUS Collaborations],
  Eur.\ Phys.\ J.\ C {\bf 75}, no. 12, 580 (2015)
  [arXiv:1506.06042 [hep-ex]].
  
\bibitem{Lai:2010vv} 
  H.~L.~Lai, M.~Guzzi, J.~Huston, Z.~Li, P.~M.~Nadolsky, J.~Pumplin and C.-P.~Yuan,
  Phys.\ Rev.\ D {\bf 82}, 074024 (2010)
  [arXiv:1007.2241 [hep-ph]].
  
\bibitem{Ball:2008by} 
  R.~D.~Ball {\it et al.} [NNPDF Collaboration],
  Nucl.\ Phys.\ B {\bf 809}, 1 (2009)
  Erratum: [Nucl.\ Phys.\ B {\bf 816}, 293 (2009)]
  [arXiv:0808.1231 [hep-ph]].
  
\bibitem{Mironov:2009uv} 
  A.~Mironov and A.~Morozov,
  JHEP {\bf 1004}, 040 (2010)
  [arXiv:0910.5670 [hep-th]].
  
\bibitem{Collins:1998rz} 
  J.~C.~Collins,
  Phys.\ Rev.\ D {\bf 58}, 094002 (1998)
  [hep-ph/9806259].
  
\bibitem{Martin:2006qz} 
  A.~D.~Martin, W.~J.~Stirling and R.~S.~Thorne,
  Phys.\ Lett.\ B {\bf 636}, 259 (2006)
  [hep-ph/0603143].
  
\bibitem{Forte:2010ta} 
  S.~Forte, E.~Laenen, P.~Nason and J.~Rojo,
  Nucl.\ Phys.\ B {\bf 834}, 116 (2010)
  [arXiv:1001.2312 [hep-ph]].
  
\bibitem{Martin:2009iq} 
  A.~D.~Martin, W.~J.~Stirling, R.~S.~Thorne and G.~Watt,
  Eur.\ Phys.\ J.\ C {\bf 63}, 189 (2009)
  [arXiv:0901.0002 [hep-ph]].
  
\bibitem{Thorne:2006qt} 
  R.~S.~Thorne,
  Phys.\ Rev.\ D {\bf 73}, 054019 (2006)
  [hep-ph/0601245].
  
\bibitem{Thorne:2012az} 
  R.~S.~Thorne,
  Phys.\ Rev.\ D {\bf 86}, 074017 (2012)
  [arXiv:1201.6180 [hep-ph]].
  
\bibitem{xFitter} 
  xFitter, An open source QCD fit framework. {\it http://xFitter.org} [xFitter.org]
  [arXiv:1410.4412 [hep-ph]].
  
\bibitem{Vafaee:2019nmo}
A.~Vafaee, K.~Javidan and A.~Shokouhi,
[arXiv:1906.07390 [hep-ph]].

\bibitem{Vafaee:2019yec}
A.~Vafaee and A.~Shokouhi,
[arXiv:1904.04285 [hep-ph]].

\bibitem{Shokouhi:2018gie}
A.~Shokouhi and A.~Vafaee,
Nucl. Part. Phys. Proc. \textbf{300-302}, 35-39 (2018)

\bibitem{Vafaee:2018ehy}
A.~Vafaee,
Nucl. Part. Phys. Proc. \textbf{300-302}, 30-34 (2018)
  
\bibitem{Vafaee:2017jnt} 
  A.~Vafaee and A.~Khorramian,
  Nucl.\ Part.\ Phys.\ Proc.\  {\bf 282-284}, 32 (2017).

  
\bibitem{Vafaee:2016jxl} 
  A.~Vafaee, A.~Khorramian, S.~Rostami and A.~Aleedaneshvar,
  Nucl.\ Part.\ Phys.\ Proc.\  {\bf 270-272}, 27 (2016).
  
  
\bibitem{Botje:2010ay} 
  M.~Botje,
  Comput.\ Phys.\ Commun.\  {\bf 182}, 490 (2011)
  [arXiv:1005.1481 [hep-ph]].
  
\bibitem{DGLAP}
 V.~N.~Gribov and L.~N.~Lipatov,
  Sov.\ J.\ Nucl.\ Phys.\  {\bf 15}, 438 (1972)
  [Yad.\ Fiz.\  {\bf 15}, 781 (1972)];\\
  L.~N.~Lipatov,
  Sov.\ J.\ Nucl.\ Phys.\  {\bf 20}, 94 (1975)
  [Yad.\ Fiz.\  {\bf 20}, 181 (1974)];\\
   Y.~L.~Dokshitzer,
  Sov.\ Phys.\ JETP {\bf 46}, 641 (1977)
  [Zh.\ Eksp.\ Teor.\ Fiz.\  {\bf 73}, 1216 (1977)];\\
G.~Altarelli and G.~Parisi,
  Nucl.\ Phys.\ B {\bf 126}, 298 (1977).   
  
\bibitem{Agashe:2014kda} 
  K.~A.~Olive {\it et al.} [Particle Data Group],
  Chin.\ Phys.\ C {\bf 38}, 090001 (2014).
   
\end{thebibliography}
\end{document}